\documentclass[%
 reprint,
 amsmath,amssymb,
 aps,
]{revtex4-2}
\usepackage{siunitx}
\usepackage{graphicx}
\usepackage{dcolumn}
\usepackage{bm}
\usepackage{color}
\usepackage{verbatim}
\usepackage[T1]{fontenc}
\usepackage[titletoc]{appendix}
\usepackage{diagbox}
\usepackage{subfigure}

\begin{document}

\preprint{APS/123-QED}

\title{Terahertz Scale Micro-Bunching Instability Driven by NEG-Coating Resistive-Wall Impedance}
\author{Weiwei Li}
\author{Tianlong He}\email{htlong@ustc.edu.cn}
\author{Zhenghe Bai}

\affiliation{National Synchrotron Radiation Laboratory, University of Science and Technology of China, Hefei, Anhui, 230029, China}%

\date{\today}

\begin{abstract}
Non-evaporable getter (NEG) coating is widely required in the next generation of light sources and circular $e^+e^-$ colliders for small vacuum pipes to improve the vacuum level, which, however, also enhances the high-frequency resistive-wall impedance and often generates a  resonator-like peak in the terahertz frequency region. In this paper, we will use the parameters of the planned  Hefei Advanced Light Facility (HALF) storage ring to study the impact of NEG coating resistive-wall impedance on the longitudinal microwave instability via particle tracking simulation. Using different NEG coating parameters (resistivity and thickness) as examples, we find that the impedance with a narrow and strong peak in the high frequency region can cause micro-bunching instability, which has a low instability threshold current and contributes to a large energy spread widening above the threshold. In order to obtain a convergent simulation of the beam dynamics, one must properly resolve such a peak. The coating with a lower resistivity has a much less sharp peak in its impedance spectrum, which is helpful to suppress the micro-bunching instability and in return contributes to a weaker microwave instability.
\begin{description}
	
\item[PACS numbers]
29.27.Bd, 41.75.Ht
\end{description}
\end{abstract}

\pacs{Valid PACS appear here}
\maketitle


\section{\label{sec1}Introduction}
The non-evaporable getter (NEG) coating \cite{Malyshev2019} has been successfully applied to inner surfaces of many vacuum chambers of particle accelerators. It can provide distributed pumping along vacuum chambers, thus, the specified ultrahigh vacuum pressure level could be met with a reduced number and a size of external pumps.

The resistivity wall (RW) impedance, produced by the finite conductivity of the beam vacuum chamber,  plays an important, often dominant, role in modern accelerators, especially in those with a small transverse size of the vacuum chamber \cite{Nagaoka2014,Blednykh2021}. The presence of the NEG coating films makes the surface resistance of the beam pipe higher than the one without coating and the resistive wall effect is more pronounced.  This impact is especially important for large machines with small beam pipe dimensions such as circular  $e^+e^-$ colliders \cite{Migliorati2018,wang2022} and diffraction-limited storage rings (DLSRs) \cite{WangNa2017,Dehler2019,Gamelin2023}.  In addition, there could be also an uncertainty on the coating conductivity measurements in the high frequency region which may give inaccurate predicts about the instability threshold. In order to reduce the RW impedance contribution and the uncertainty due to the mostly unknown coating resistivity,  one of the best ways is to reduce the NEG coating thickness. Several DLSR projects \cite{WangNa2017,Dehler2019,Gamelin2023}  have set the NEG coating thickness targets equal to or less than 1 um. Recent study \cite{Gamelin2023} also shows that there is a regime in which a coating with a lower resistivity produces even a larger loss factor than that with a higher resistivity.

\begin{table*}
\caption{\label{tab:wall}Main Parameters of the Vacuum Chambers}
\begin{ruledtabular}
\begin{tabular}{ccccc}
 Type & Material + film (thickness: $\si{\micro\metre}$)  &Shape & Aperture/Radii (mm)& Length (m)    \\
        \hline
           Main Chamber  & CuCrZr+NEG (d) &  Round & 13 &344.2        \\
           Fast Corrector & Inconel + NEG (d) & Round & 13 & 7.2 \\
           beam pipes with antechamber &  Stainless steel + Cu (20) & Round & 13 & 48.2 \\
           out vacuum Insertion devices & Al &  Elliptical &   $8\times26$   & 43 \\
           In-Vacuum Undulators & NdFeB + Ni (75) + Cu (75) & Rectangular & $6\times65$ & 5.4 \\
           Others (ie., bellows, flanges) & Stainless Steel  & Round  & 13   & 32
\end{tabular}
\end{ruledtabular}
\end{table*}

The RW  has a strong longitudinal impedance in the high frequency region, contributing to a sharp variation of the point-charge longitudinal wakefield in very short distances. The presence of the NEG coating will further enhance the high-frequency impedance and often generates a resonator-like peak \cite{Shobuda2017,Motlagh2023}. The multi-particle tracking  simulations \cite{PyHEADTAIL,SKRIPKA2016,Xu2019,Wang2023} are widely used to study the beam dynamics but may have computational issues to study the longitudinal microwave instability (MWI), where a large number of simulation particles is needed to study the response of small-scale bunch structures to high frequency wakefield components. If not equipped with suitable algorithms (i.e., smoothing/filtering techniques, fine grids, etc.), the simulation can fail to produce reliable results \cite{Bassi2016}. Since the collective behavior usually doesn't depend on the behavior of the wakefield/impedance at very small length scales/very high frequencies, one popular solution is to use the wake potential of a very short Gaussian bunch of rms length $\overline{\sigma}_s$ (sometimes called pseudo-Green function) in place of the point-charge wakefield \cite{Blednykh2021,Migliorati2018,WANGDAN2022,Carver2023}. Then the impedance used in tracking simulation becomes that of the point-charge multiplied  by a Gaussian filter. The remaining item is to determine the required length $\overline{\sigma}_s$, which effectively means finding the frequency range over which the impedance affects the dynamics. Usually the guideline that  is 10 or 15 times smaller than that of  the equilibrium beam gives a reasonable estimate \cite{Blednykh2021} . However, the MWI simulations are quite sensitive to numerical noise, so it is necessary to vary different tracking parameters to make sure that the tracking results are accurate.

In this paper, the Hefei Advanced Light Facility (HALF) \cite{Bai2021}, a fourth generation light source in design, will be used to study the impact of the NEG coating parameters. The equilibrium beam  rms length at zero current limit is 2.1 mm, however, in order to obtain a quasi convergent simulation of the beam dynamics, the required length $\overline{\sigma}_s$ should typically be even smaller than 0.02 mm and millions of marco-particles are necessary. Under convergent simulations, we find that the coating with a high resistivity (in the order of $10^{-5}~\si{\Omega~m} $), whose  impedance has a sharp peak in the terahertz region, can cause an undesirable micro-bunching instability (MBI), with a low threshold and a large energy spread widening.

     This paper is organized as follows. In Sec. II, the RW impedance with different NEG coating parameters of HALF will be presented. In Sec. III, the particle tracking method will be  introduced. Section IV shows the simulation results with different tracking and coating parameters. The conclusions and discussions are presented in Sec. V.

\section{\label{sec2} resistive wall impedance in half}

We consider a simplified model of the ring consisting 6 parts and the main parameters of the vacuum chambers are listed in Table ~\ref{tab:wall}. The resistivities of the materials are listed in Table ~\ref{tab:material}.

\begin{table}[!hbt]
\setlength{\tabcolsep}{3.5mm}
   \centering
   \caption{Main parameters of HALF}
   \begin{tabular}{cc}
       \toprule
       Material                   & Resistivity ($\si{\Omega~m} $)             \\
       \hline
          Cu & $1.68\times 10^{-8}$ \\
          CuCrZr & $2.3\times 10^{-8}$ \\
          Al6063 & $3.16\times 10^{-8}$ \\
          Ni &  $6.93\times 10^{-7}$ \\
          SS316L & $7.41\times 10^{-7}$ \\
          Inconel 625 & $1.29\times 10^{-6}$ \\
       \hline
   \end{tabular}
   \label{tab:material}
\end{table}

 The NEG resistivity value depends on the compound composition and coating method \cite{MALYSHEV2017}. The resistivity measurement results also have very large discrepancy using different methods \cite{Plouviez2018}. Thus three different  resistivity $\rho_{\si{NEG}}$ values will be taken into account: $1\times 10^{-5}~\si{\Omega~m}$, $5\times 10^{-6}~\si{\Omega~m}$ and $1\times 10^{-6}~\si{\Omega~m}$.
 The target of NEG coating thickness is $d=1~\si{\micro\metre}$ for the HALF project, but two different film thicknesses  will be studied for comparison: 0.5  $\si{\micro\metre}$ and 1  $\si{\micro\metre}$. 

\begin{figure}
\centering\includegraphics[width=8.5cm]{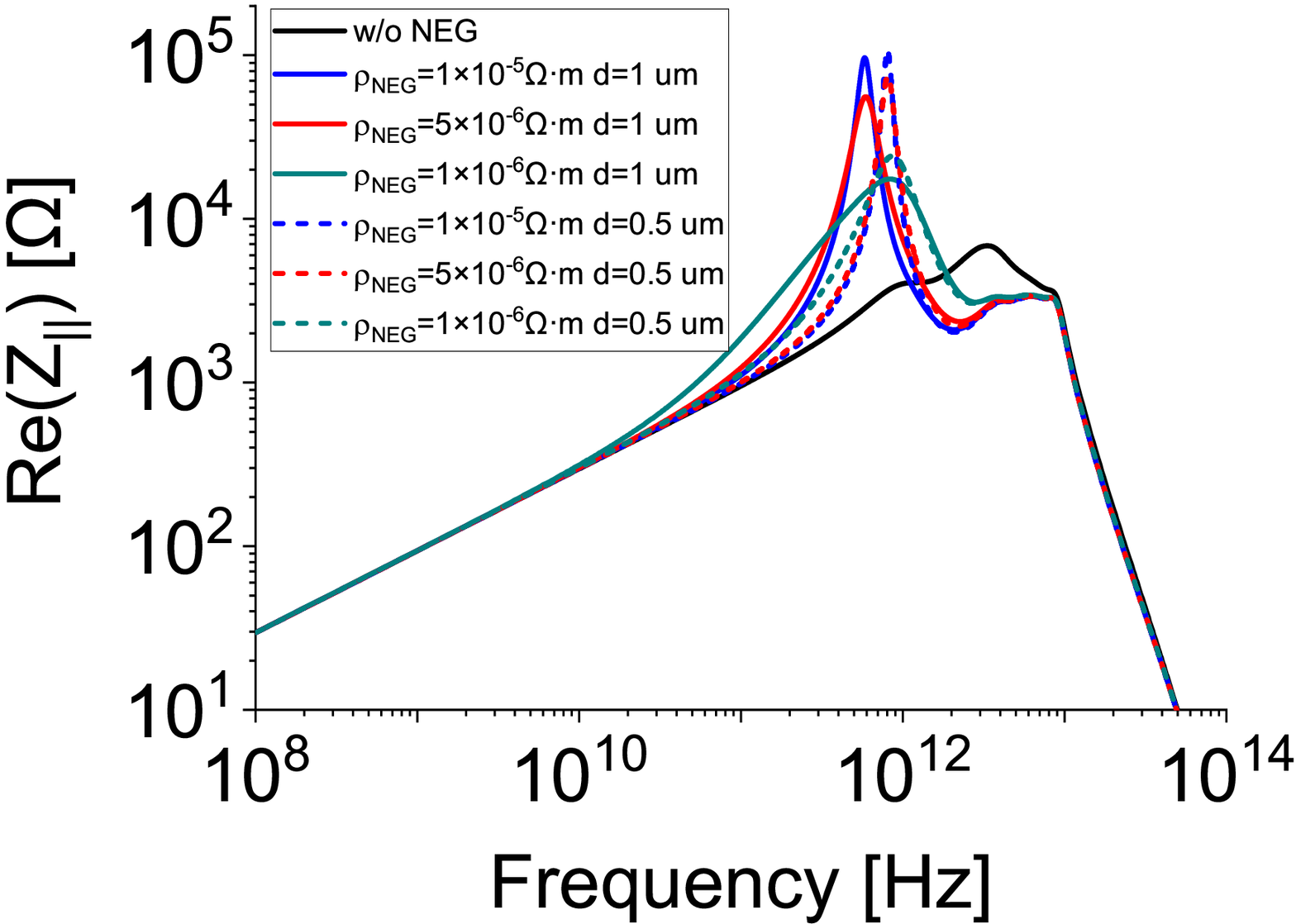}
\centering\includegraphics[width=8.5cm]{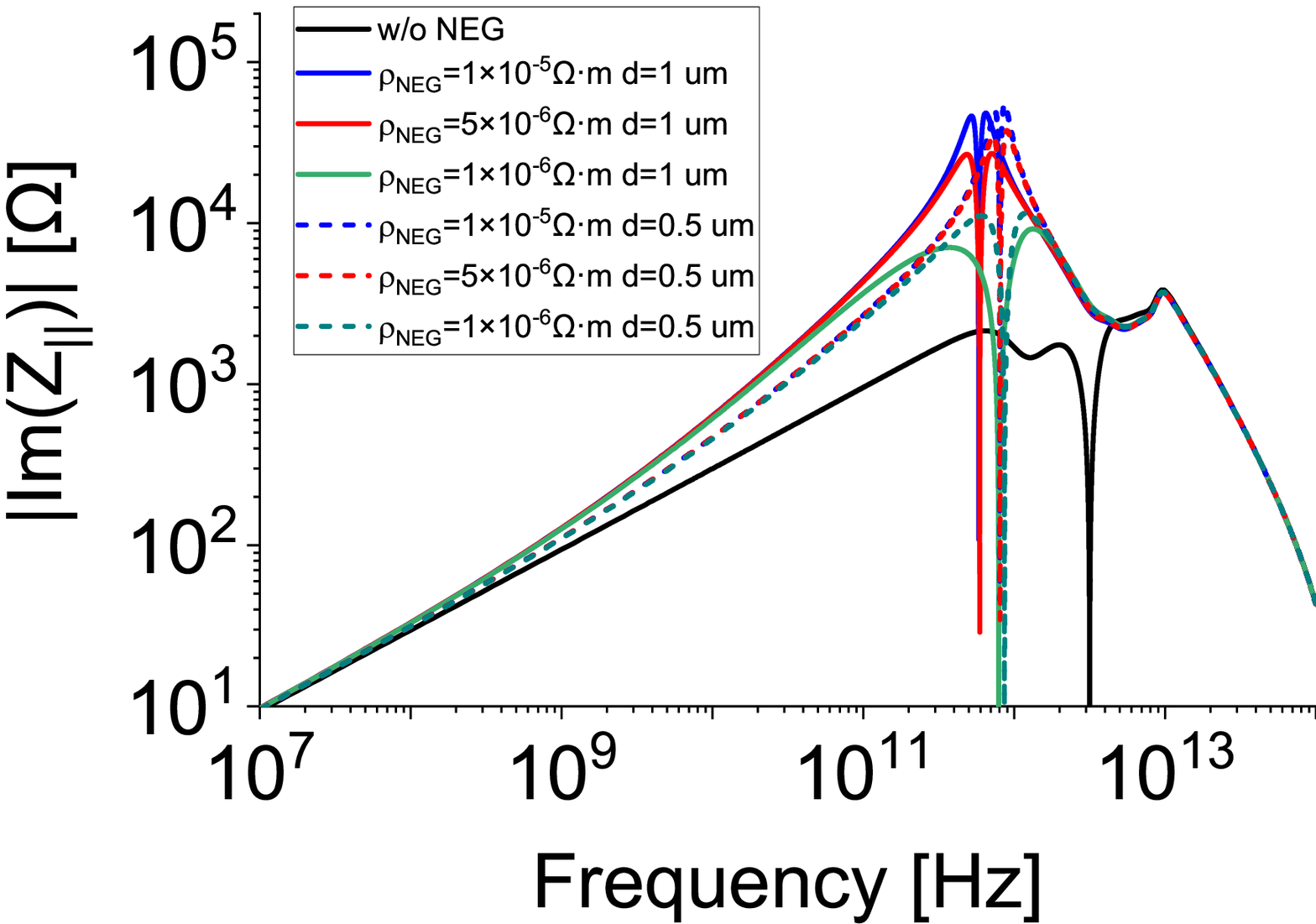}
\caption{The real (top) and imagine (bottom) parts of the longitudinal impedance for different coating parameters.}
\label{fig1}
\end{figure}

The resistive wall impedance is computed using the ImpedanceWake2D (IW2D) code \cite{IW2D}, which solves for a  circular geometry and then applies a Yokoya factor for elliptical or rectangular cases \cite{Yokoya1993,Migliorati2019}. Their longitudinal impedance as a function of frequency is shown in Fig. \ref{fig1}.  At low frequency, all the impedance is similar since the thickness of coating is much smaller than its skin depth. At high frequency, the impedance is greatly enhanced with NEG coatings and generates a resonator-like peak in the terahertz region.  If the coating resistivity or thickness is smaller, the impedance will be more close to that of no coating. A smaller coating resistivity will reduce the quality factor and the peak impedance. A thinner coating thickness will make the peak frequency shift downward.

It is useful to define the effective impedance as \cite{Chao1993,Ng2005,SMALUK2018}
\begin{equation}\label{eql}
{\left( {\frac{{{Z_\parallel }}}{n}} \right)_{{\rm{eff}}}} = \frac{{\int_{ - \infty }^\infty  {{Z_\parallel }\left( \omega  \right)\frac{{{\omega _0}}}{\omega }h\left( \omega  \right)d\omega } }}{{\int_{ - \infty }^\infty  {h\left( \omega  \right)d\omega } }},
\end{equation}
where $n=\omega/\omega_0$ is the revolution harmonic number, $\omega_0$ is the revoltion angular frequency, $h\left( \omega  \right) = \tilde \lambda \left( \omega  \right){{\tilde \lambda }^*}\left( \omega  \right)$ is the bunch power spectrum, $\tilde\lambda \left( \omega  \right)$ is the Fourier transform of the longitudinal charge density $\lambda \left(t\right)$. Assuming a Gaussian bunch, $h\left( \omega  \right)=e^{-\omega^2\sigma_t^2}$, where $\sigma_t$ is the rms bunch length in time. The effective impedance with natural bunch length (7 ps) for different NEG coating parameters are listed in Table  ~\ref{tab:imp}.

\begin{table}[!hbt]
   \centering
   \caption{The effective impedance  $(\si{m\Omega})$  with natural bunch length for different NEG coating parameters}
   \begin{tabular}{|c|c|c|c|}
     \hline
     \diagbox{d ($\si{\micro\metre}$)} {$\rho_{\si{NEG}} (\si{\Omega~m})$ }  &    $1\times 10^{-5}$& $5\times 10^{-6}$ & $1\times 10^{-6}$          \\
       \hline
          1 & 77.7 & 77.5 & 76.5 \\
          \hline
          0.5 & 67.2 & 67.1 & 66.6 \\
       \hline
   \end{tabular}
   \label{tab:imp}
\end{table}

\section{\label{sec3} Multi-particle Tracking Simulation}

 Most of the macroparticle tracking codes compute the wake potential as the convolution between the longitudinal wake function $w_\parallel(t)$, i.e., the Green function of a point charge, and the bunch distribution  $\lambda \left(t\right)$:  \cite{SKRIPKA2016}
\begin{equation}\label{eq2}
{W_\parallel }\left( t \right) = \int_{ - \infty }^t {{w_\parallel }\left( {t - t'} \right)\lambda \left( {t'} \right)} dt'. 
\end{equation}
The longitudinal wake function can be expressed in terms of the longitudinal impedance by an inverse Fourier transform
\begin{equation}\label{eq3}
{w_\parallel }\left( t \right) = \frac{1}{{2\pi }}\int_{ - \infty }^\infty  {{Z_\parallel }\left( \omega  \right){e^{i\omega t}}} d\omega. 
\end{equation}

For the convenience of numerical calculation, the time coordinate in Eq.~\ref{eq2} should be equi-spaced and discrete Fourier Transforms can be used. The bin size is $\Delta_t=\frac{0.5}{F_m}$, where $F_m$ is the maximum frequency of the impedance in Eq.~\ref{eq3}.  $\lambda \left(t\right)$ is obtained by counting the particle number in each bin and  $ > \sim$ 1000 particles per bin is typically required. However, the input RW wake function can cause some problems  to simulations based on this approach, since it covers a very high frequency range and a very large number of slices would be necessary, thus increasing the computational load. As discussed in Sec.~\ref{sec1}, the pseudo-Green function from a very short Gaussian pulse will be used. The pulse length $\overline{\sigma}_s$ will determine the frequency reach of the impedance calculation and needs to be much smaller than the real bunch length used in tracking simulations to cover the spectrum of interest. However, in order to resolve the  impedance peak, the computational load is still very heavy.

The STABLE code \cite{STABLE} will be dedicated to conduct multi-particle tracking simulations for longitudinal beam dynamics studies. 
It is implemented in a MATLAB environment with the usage of the state-of-the-art of graphics-processing-unit acceleration technique so that the tracking efficiency is significantly improved. The original version of STABLE is written for multi-bunch and multi particle simulation, and a 2D matrix is used to store the macro-particle's coordinates, with each column corresponding to one bunch. In order to accurately simulate the single bunch dynamics, which usually requires millions or even tens of millions of macro-particles, we need only modify the STABLE code by dividing the macro-particles of single bunch into multiple parts and storing them in each column of the 2D matrix. We can separately count the bin-distribution of each column, and then summarize them to obtain the total bunch distribution. In addition, a fixed bin width instead of the number of bins is set in default. Therefore, the bin number will increase as the bunch lengthening.  The remaining operations, such as convolution of the bunch distribution and the short-range wake (or short-bunch wakepotential), and interpolation to obtain the short-range wake kick of each macro-particle, can be same as those in the original version of STABLE. 

\section{\label{sec4} NUMERICAL RESULTS FOR THE PARAMETERS OF HALF}

 In this section, the impact of NEG-coated resistive-wall impedance  on the longitudinal beam dynamics is investigated by tracking simulations in the framework of HALF project. The main parameters of the HALF storage ring with insertion devices are summarized in Table~\ref{tab:HALF}. 

 \begin{table}[!hbt]
\setlength{\tabcolsep}{3.5mm}
   \centering
   \caption{Main parameters of HALF}
   \begin{tabular}{lcc}
       \toprule
       \textbf{Parameter}&{Symbol}                  &{Value}               \\
       \hline
           Ring circumference  & C           & \SI{480}{m}         \\
           Beam energy & $E_0$              & \SI{2.2}{GeV}          \\
           Nominal beam current & $I_0$ &\SI{350}{mA}         \\
           Longitudinal damping time & $\tau_z$       & \SI{14}{ms}      \\
           Momentum compaction &
           $\alpha_c$
           & $9.4\times10^{-5}$           \\
           Natural energy spread  &
           $\sigma_\delta$ &$7.3\times10^{-4}$           \\
           Harmonic number & $h$          & \SI{800}{}         \\
           Energy loss per turn &
           $U_0$
           & \SI{400}{keV}         \\
           Voltage of MC &
           $V_{RF}$
           & \SI{1.2}{MV}        \\
           Natural rms bunch length & $\sigma_{t0}$              & \SI{7}{ps}        \\
       \hline
   \end{tabular}
   \label{tab:HALF}
\end{table}
 
\subsection{\label{sec4a}Convergence Study}
We  first carry out the convergence study with two typical examples, the coatings with $\rho_{\si{NEG}} =1\times10^{-5}~\si{\Omega~m}$ $d=1~\si{\micro\metre} $ and  $\rho_{\si{NEG}} =1\times10^{-6}~\si{\Omega~m}$, $d=1~\si{\micro\metre} $   respectively. 

\subsubsection{\label{sec4a1}Pseudo-Green Function}
\begin{figure}
\centering\includegraphics[width=8.5cm]{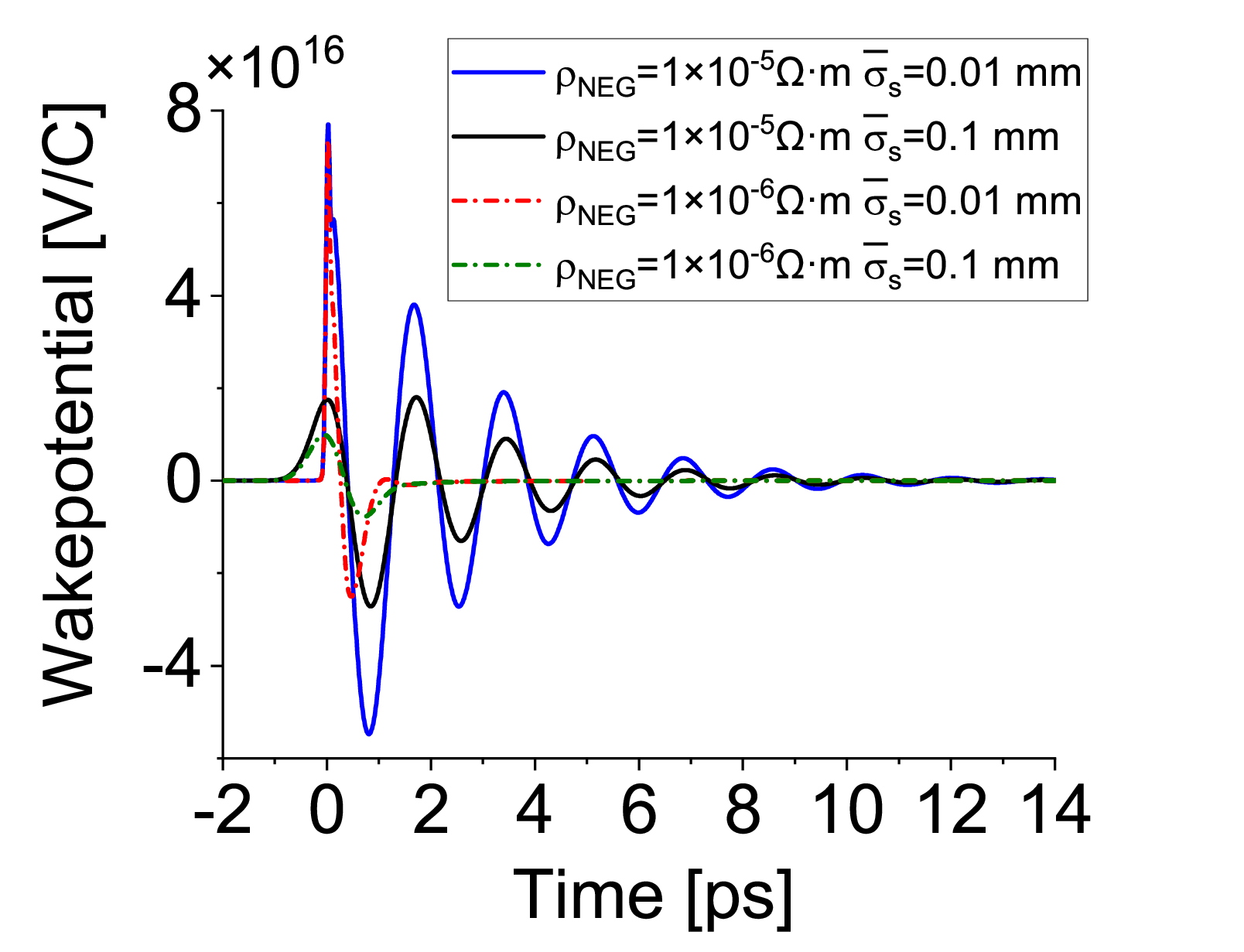}
\caption{Longitudinal wake potentials for different $\rho_{\si{NEG}}$ and $\overline{\sigma}_s$. The coating thickness is $d=1~\si{\micro\metre} $.}
\label{fig2}
\end{figure}

The wake potentials for the short Gaussian bunch of  $\overline{\sigma}_s=$ 0.1 mm and 0.01 mm are shown in Fig.~\ref{fig2}, where the positive (negative) value means energy loss (gain), and they will be used instead of the wake function generated from a point charge. To calculate the effective impedance or loss factors for a perfect Gaussian bunch distribution with nominal bunch length $\sigma_{t0}$ from the pseudo-Green function, using the wake potentials of $\overline{\sigma}_s=$ 0.1 mm can usually obtain sufficiently accurate results. Note that $\overline{\sigma}_s=$ 0.1 mm is more than 20 times shorter than the natural bunch length $\sigma_{t0}$.

\begin{figure}
\centering\includegraphics[width=8.5cm]{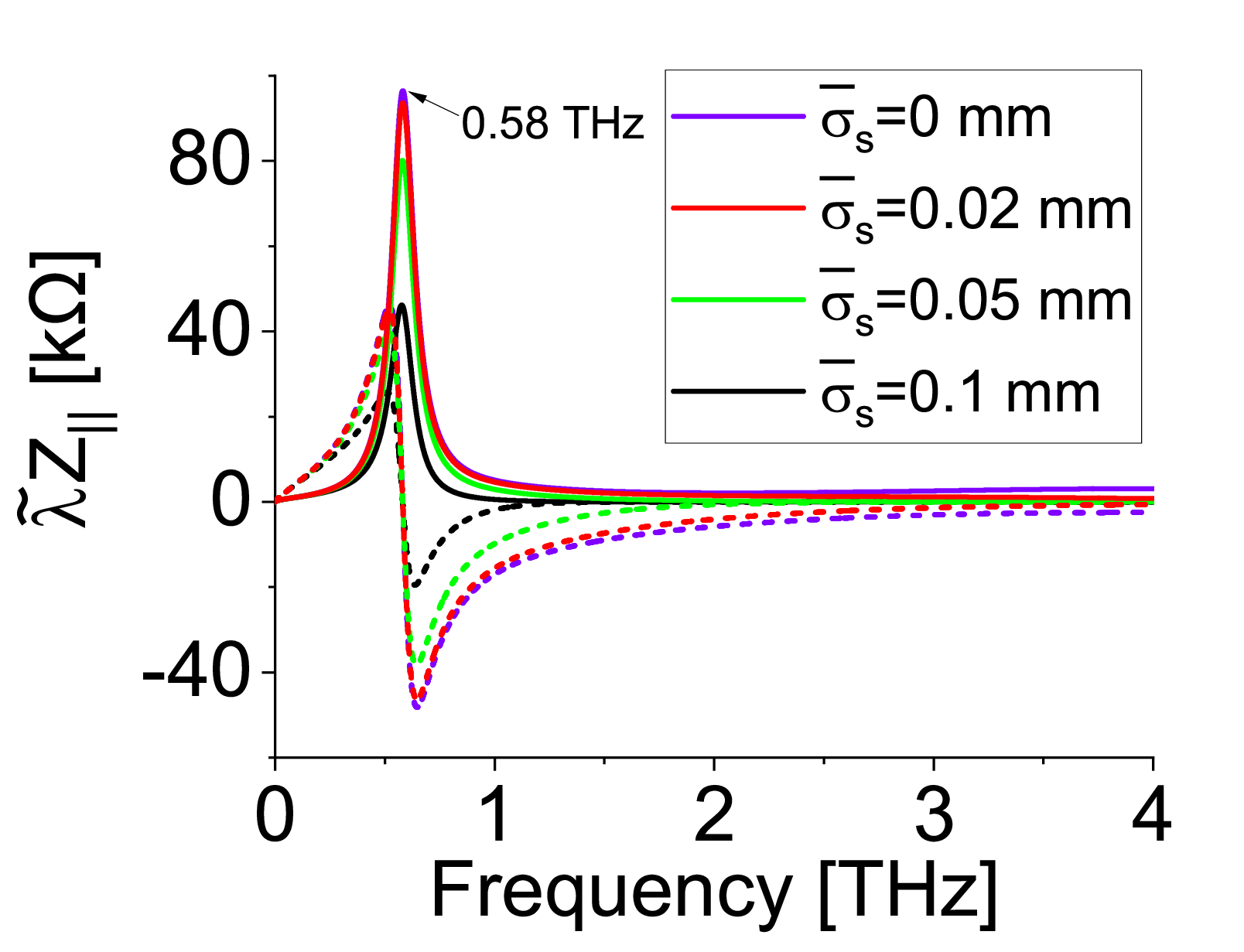}
\centering\includegraphics[width=8.5cm]{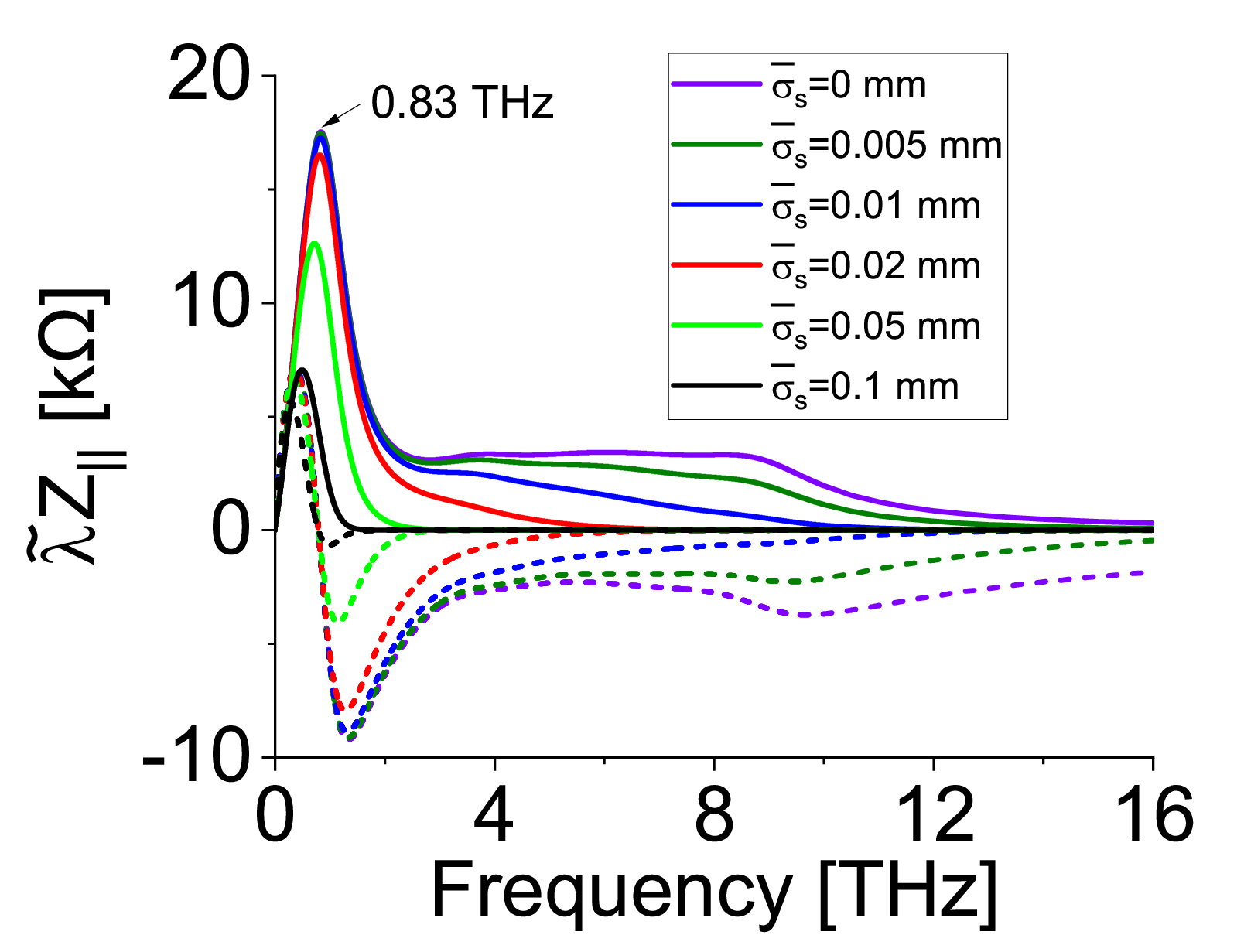}
\caption{ The real part (solid) and the absolute value of the imagine part (dashed) of the longitudinal impedance  multiplied by different Gaussian filters for $\rho_{\si{NEG}} =1\times10^{-5}~\si{\Omega~m}$ (top) and $\rho_{\si{NEG}} =1\times10^{-6}~\si{\Omega~m}$ (bottom).The coating thickness is $d=1~\si{\micro\metre} $.}
\label{fig3}
\end{figure}

The longitudinal impedance multiplied by the Fourier spectrum $\tilde\lambda \left( \omega  \right)$ of the Gaussian bunches with different  $\overline{\sigma}_s$    is plotted in Fig.~\ref{fig3}. In order to resolve the impedance peak clearly, $\overline{\sigma}_s$ should be smaller than 0.02 mm.  

\subsubsection{\label{sec4a2}The Case for $\rho_{\si{NEG}} =1\times10^{-5}~\si{\Omega~m}$ and $d=1~\si{\micro\metre}$ }

\begin{figure}
\centering\includegraphics[width=8.5cm]{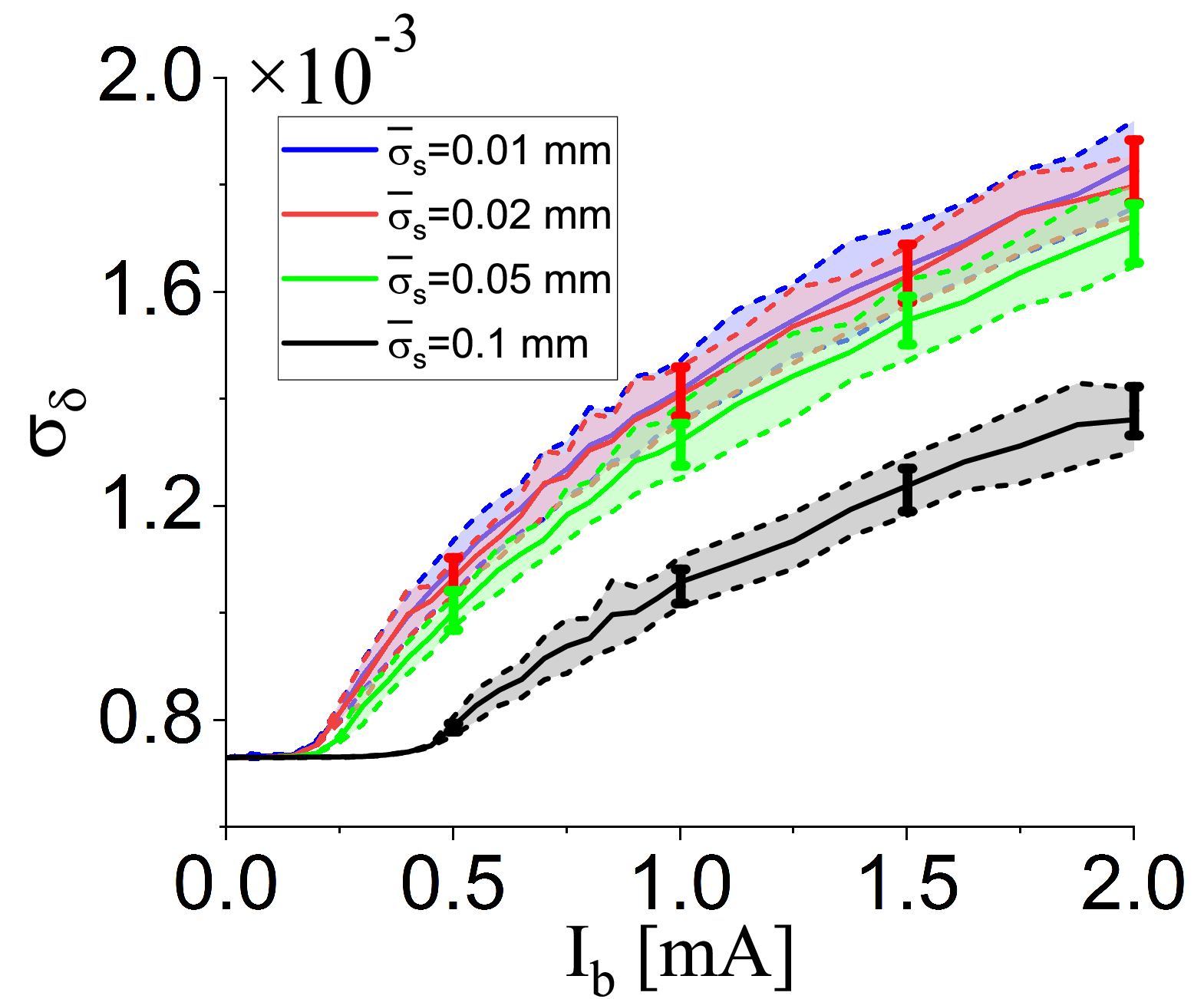}
\centering\includegraphics[width=8.7cm]{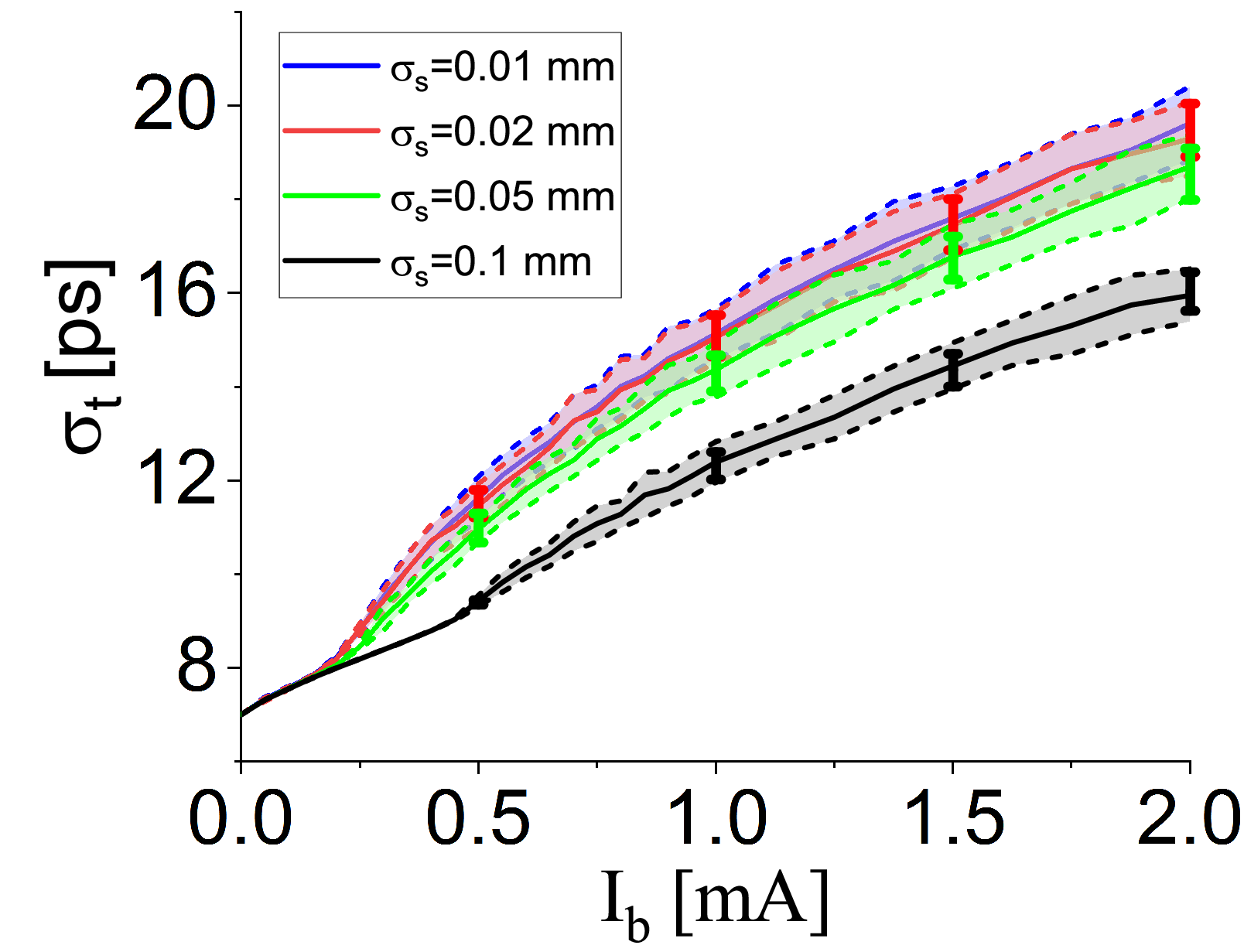}
\caption{ The rms energy spread (top) and bunch length (bottom) versus single bunch current for $\rho_{\si{NEG}} =1\times10^{-5}~\si{\Omega~m}$ and $d=1~\si{\micro\metre}$  with different $\overline{\sigma}_s$. The solid lines are the mean values and the dashed lines including their fill areas represent the standard deviation obtained by STABLE with current step of 0.05 mA. The discrete error bars are obtained by Pelegant with current step of 0.5 mA.}
\label{fig4}
\end{figure}

Figure ~\ref{fig4} shows the predicted energy spread and bunch lengthening from the tracking simulations as a function of the single bunch current for $\rho_{\si{NEG}} =1\times10^{-5}~\si{\Omega~m}$ and $d=1~\si{\micro\metre}$ with various values of $\overline{\sigma}_s$, where the particle number $N_p$ is 5 millions (M), 40,000 turns are tracked for each current and the bin size $\Delta_t$ is set to be 0.02 ps. The corresponding $F_m$ is 25 THz, which should be high enough to cover the frequency region of interest. Another guideline is that $\Delta_t$ should be small enough to resolve the wake potential generated from the short Gaussian bunch with  $\overline{\sigma}_s$. The mean value and standard deviation of bunch length and relative energy spread are computed on last 10,000 turns.  To benchmark the simulation results from the STBALE code, the Pelegant code \cite{ELEGANT}  is also used with the same parameters except for fewer current and $\overline{\sigma}_s$ values and the results are also marked in Fig.~\ref{fig4}. In order to obtain one data point in Fig.~\ref{fig4}, it takes about 200 min for Pelegant using 80 CPU cores, while less than 10 min for STABLE using 3584 CUDA cores. Good agreements are achieved since their underlying physical models are the same. Thus we will only use the STABLE code for the other simulations. 

\begin{figure}
\centering\includegraphics[width=8.5cm]{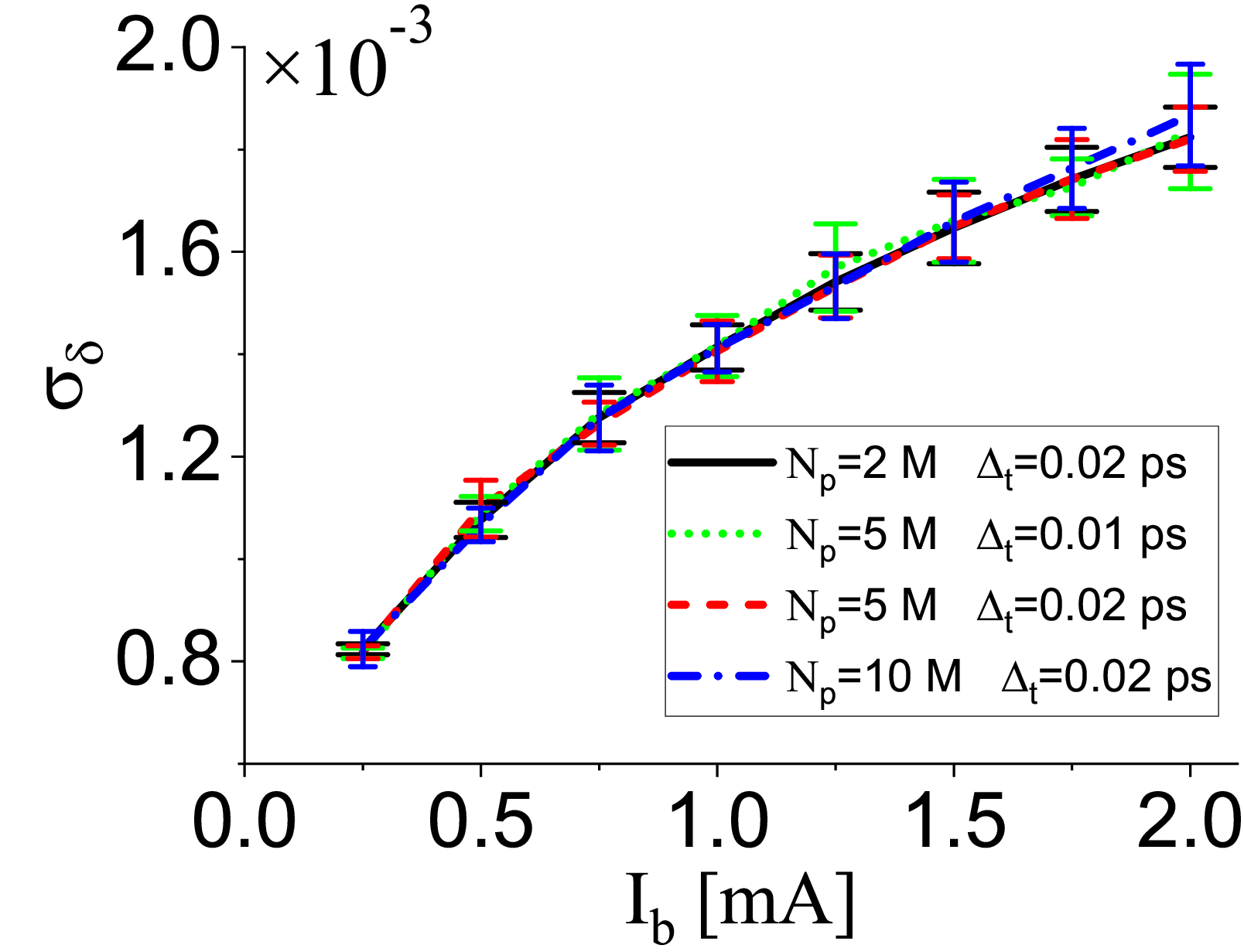}
\caption{The energy spread  versus single bunch current for $\rho_{\si{NEG}} =1\times10^{-5}~\si{\Omega~m}$ and $d=1~\si{\micro\metre}$  $\overline{\sigma}_s$=0.01 mm with different $N_p$ and $\Delta_t$. The  lines represent the mean values and the error bars represent the standard deviation.}
\label{fig5}
\end{figure}

To validate the choice of the bin size $\Delta_t$ and the particle number $N_p$, convergence studies are performed  for the the case of $\overline{\sigma}_s=$ 0.01 mm, since a shorter $\overline{\sigma}_s$ requires  more severe convergence conditions. The corresponding energy spreads are shown in Fig.~\ref{fig5}. There is no significant variation when  $\Delta_t$ decreases from 0.02 ps to 0.01 ps or  $N_p$ varies from 2 M to 10 M.

As seen in  Fig.~\ref{fig4}, at the low current below the MWI threshold, the simulations using a relative long $\overline{\sigma}_s=$ 0.01 mm  have already given enough convergent results.  However, in order to accurately evaluate the MWI, one must properly resolve the resonator-like peak impedance. To obtain a full convergent simulation for the coating with $\rho_{\si{NEG}} =1\times10^{-5}~\si{\Omega~m}$ and $d=1~\si{\micro\metre}$, the required $\overline{\sigma}_s$ is 0.02 mm. The MWI behavior can be significantly underestimated if the wakefield resolution $\overline{\sigma}_s$ is not sufficient.

\subsubsection{\label{sec4a3}The Case for $\rho_{\si{NEG}} =1\times10^{-6}~\si{\Omega~m}$ and $d=1~\si{\micro\metre}$ }

\begin{figure}
\centering\includegraphics[width=8.5cm]{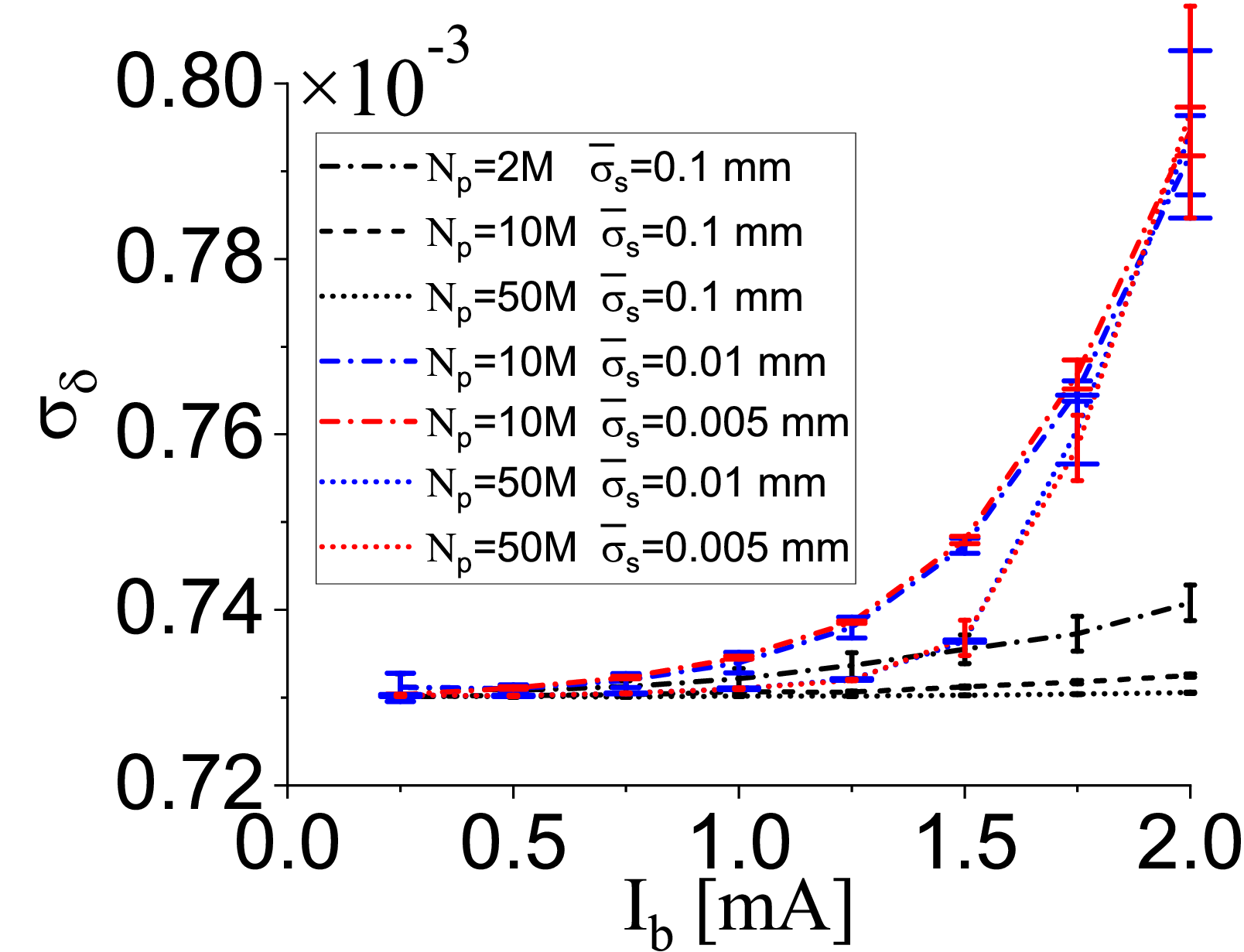}
\caption{The rms energy spread  versus single bunch current for $\rho_{\si{NEG}} =1\times10^{-6}~\si{\Omega~m}$ and $d=1~\si{\micro\metre}$   with different $N_p$ and $\overline{\sigma}_s$. The  lines represent the mean values and the error bars represent the standard deviation.}
\label{fig6}
\end{figure}

Figure~\ref{fig6} shows the predicted energy spread from the tracking simulations as a function of the single bunch current for $\rho_{\si{NEG}} =1\times10^{-6}~\si{\Omega~m}$ and $d=1~\si{\micro\metre}$  with various values of $\overline{\sigma}_s$  and $N_p$, where $\Delta_t=$ 0.02 ps and the current step is 0.25 mA. For $\overline{\sigma}_s=$ 0.1 mm, if $N_p=$ 2 M is adopted, it shows obvious energy spread widening, but  as with the increment of $N_p$, the energy spread widening becomes smaller and when  $N_p$ increases to 50 M, there is nearly no MWI within 2 mA. However, the peak impedance is still not resolved, thus we further study the case of  $\overline{\sigma}_s=$ 0.01 mm and 0.005 mm. With the same $N_p$, their results are close, so $\overline{\sigma}_s=$ 0.01 mm should be enough to cover the frequency region of interest. Within 1.5 mA, the energy spread widening becomes smaller as with the increment of $N_p$, but the full convergence is still not achieved even when $N_p=$ 50 M, but the energy spread widening becomes relative small. Therefore we can conclude that there is no or very weak MWI within 1.5 mA and it is reasonable to use a long $\overline{\sigma}_s$ such as 0.1 mm to filter out the high frequency wakefield components. For $\overline{\sigma}_s=$ 0.01 mm or 0.005 mm at a high current of 2 mA,  there is no significant variation when $N_p$ varies from 10 M to 50 M and obvious energy spread widening can be seen, so in this situation one should also use a small $\overline{\sigma}_s$ to resolve the resonator-like peak impedance in order to accurately predict the MWI behavior. 

\subsection{\label{sec4b}Micro-bunching Instability Phenomena}

In the previous subsection, it has shown that the coating with  $\rho_{\si{NEG}} =1\times10^{-5}~\si{\Omega~m}$ and $d=1~\si{\micro\metre}$ causes a much more serious MWI than that with $\rho_{\si{NEG}} =1\times10^{-6}~\si{\Omega~m}$ and $d=1~\si{\micro\metre}$ although they are close in effective impedance with natural bunch length, so there is a practical interest in exploring the underlying mechanism.

\begin{figure}
\centering\includegraphics[width=8.5cm]{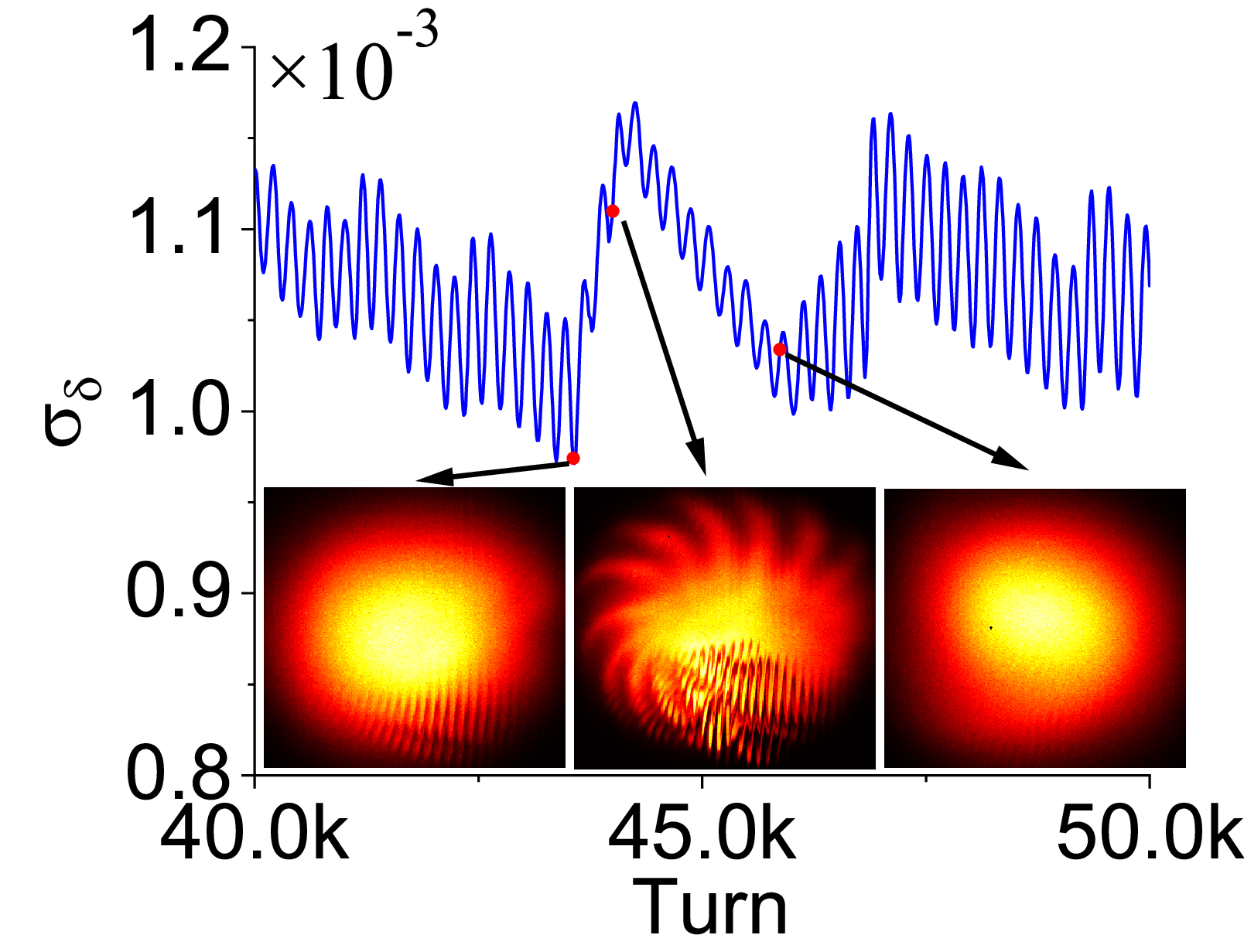}
\caption{The rms energy spread evolution over the pass turn for $\rho_{\si{NEG}} =1\times10^{-5}~\si{\Omega~m}$ and $d=1~\si{\micro\metre}$  at 0.5 mA together with the longitudinal phase space plots ($t-\delta$) at three marked points.}
\label{fig7}
\end{figure}

The energy spread evolution over the pass turn for $\rho_{\si{NEG}} =1\times10^{-5}~\si{\Omega~m}$ and $d=1~\si{\micro\metre}$ with $\overline{\sigma}_s=$ 0.01 mm and $N_p=$ 20 M at current of 0.5 mA  together with the longitudinal phase space distributions at three different turns are shown in Fig.~\ref{fig7}. More simulated particles are used just to make the plots of the phase spaces more smooth and clear. There appears strong sawtooth-shaped fluctuations of energy spread over the pass turn. The error bars in Figs.~\ref{fig4} and \ref{fig5} also characterize the amplitudes of the fluctuations. The MBI in the phase space is visible corresponding to a modulation frequency around 0.58 THz. It implies that the sharp peak in the impedance spectrum plays an important role in the MBI. A possible reason is that  when the peak is narrowband (or has a high quality factor), the corresponding wakefield lasts for several oscillation cycles (as shown in Fig.~\ref{fig2}), which allows the wakefield from the micro-bunches far apart to be coherently enhanced.  While for $\rho_{\si{NEG}} =1\times10^{-6}~\si{\Omega~m}$ and $d=1~\si{\micro\metre}$, the peak is more broadband, the wakefield attenuates quickly with the increasing of distance, which prevents the cooperation between the micro-bunching fluctuations far apart.      

\subsection{\label{secc}Impact of Coating Parameters}

We have shown the coating with  $\rho_{\si{NEG}} =1\times10^{-5}~\si{\Omega~m}$ and $d=1~\si{\micro\metre}$ can cause the MBI effect with a low current threshold for the HALF ring. To avoid its occurrence, there is a practical interest in exploring the dependence on  the coating parameters. 

\begin{figure}
\centering\includegraphics[width=8.5cm]{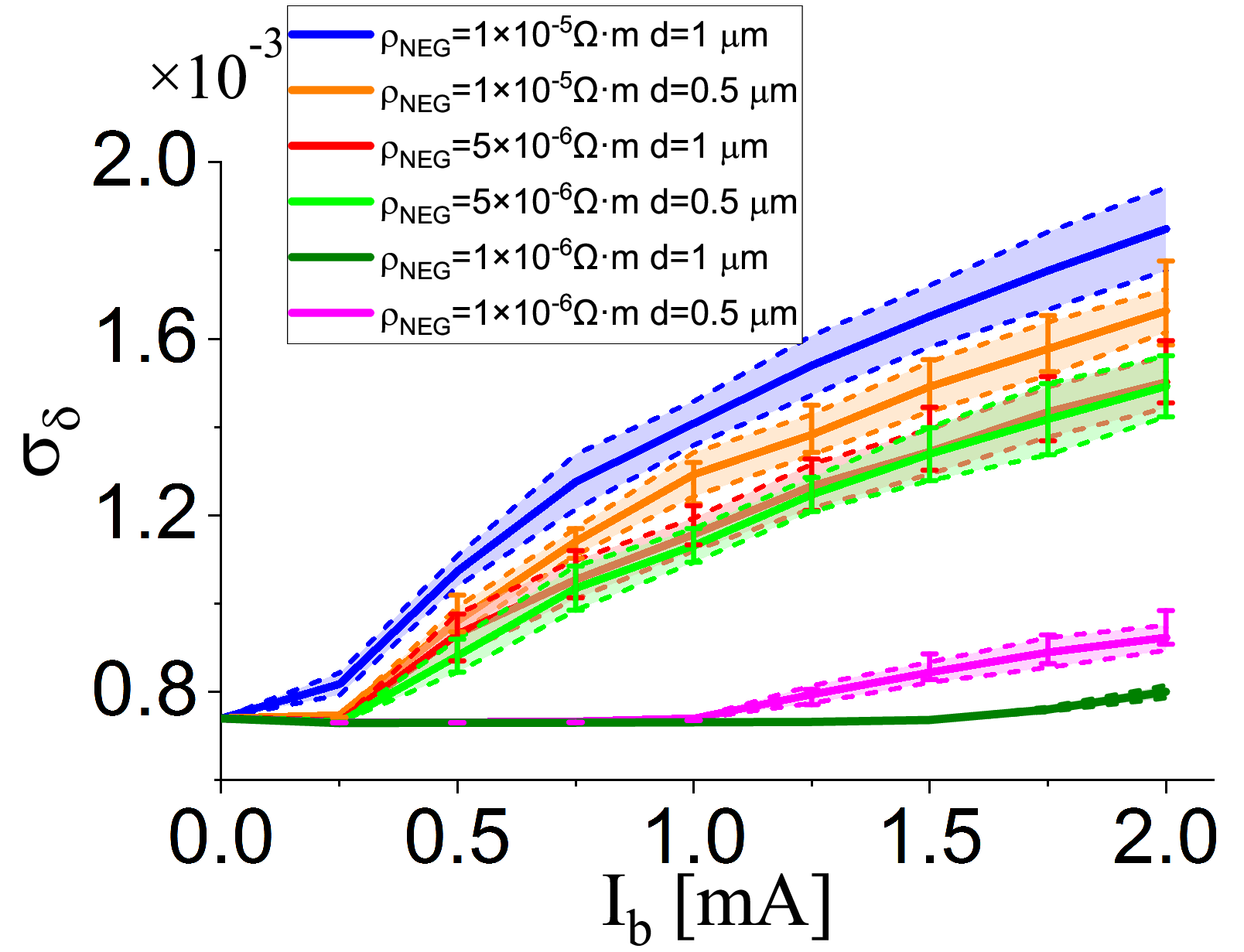}
\centering\includegraphics[width=8.5cm]{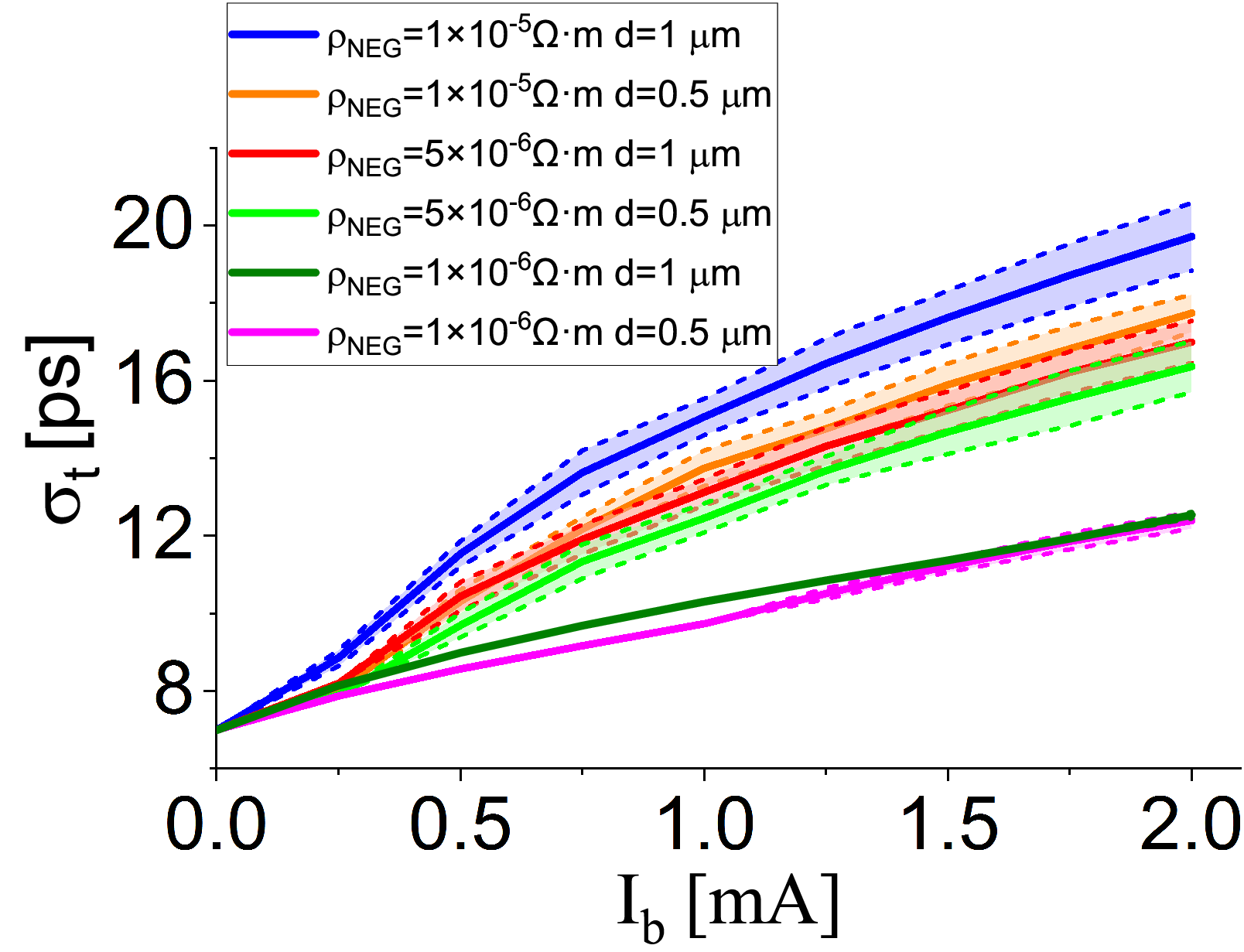}
\caption{The rms energy spread (top) and bunch length (bottom) versus single bunch current for different coatings. The solid lines are the mean values and the dashed lines including their fill areas represent the standard deviation. The discrete error bars are obtained using $N_p=$ 20 M to validate the convergences.}
\label{fig8}
\end{figure}

We consider the coating parameters given in Sec.~\ref{sec2}, in which their impedances are given.  Figure~\ref{fig8} shows the predicted energy spread and bunch lengthening from the tracking simulations as a function of the single bunch current for different coating parameters, where the current step is 0.25 mA, $\Delta_t=$ 0.02 ps,  $\overline{\sigma}_s=$ 0.01 mm which is small enough to resolve the resonator-like peak impedance, and the particle number is 10 M except for the case of $\rho_{\si{NEG}} =1\times10^{-6}~\si{\Omega~m}$ and $d=1~\si{\micro\metre}$ where 50 M particles are used. The convergence studies for the cases $\rho_{\si{NEG}} =1\times10^{-5}~\si{\Omega~m}$, $d=1~\si{\micro\metre}$ and $\rho_{\si{NEG}} =1\times10^{-6}~\si{\Omega~m}$, $d=1~\si{\micro\metre}$ have been done in the subsection \ref{sec4a}. To validate the choice of $N_p$ for the other cases, we also carry out the simulations using $N_p=$ 20 M and the results are also plotted in Fig.~\ref{fig8} as error bars, and  there is no significant variation for each case. 
 With the same coating thickness $d=1~\si{\micro\metre}$ or $0.5~\si{\micro\metre}$, the MWI for $\rho_{\si{NEG}} =5\times10^{-6}~\si{\Omega~m}$ is less serious and has a higher threshold current than that for $\rho_{\si{NEG}} =1\times10^{-5}~\si{\Omega~m}$, but there still exits MBI  when  the current exceeds the threshold. For the coating with resistivity $\rho_{\si{NEG}} =1\times10^{-5}~\si{\Omega~m}$ or  $5\times10^{-6}~\si{\Omega~m}$, reducing the thickness from $1~\si{\micro\metre}$ down to $0.5~\si{\micro\metre}$ is helpful to weaken the MBI, since the peak frequency in the impedance spectrum for $d=0.5~\si{\micro\metre}$ is higher than that for $d=1~\si{\micro\metre}$ as seen in Fig.~\ref{fig1}.   For the coating with $\rho_{\si{NEG}} =1\times10^{-6}~\si{\Omega~m}$ and $d=1~\si{\micro\metre}$, it doesn’t  contribute to MBI, so it has a much higher instability threshold and smaller energy spread widening.  For the coating with $\rho_{\si{NEG}} =1\times10^{-6}~\si{\Omega~m}$, reducing the coating thickness from $1~\si{\micro\metre}$ down to $0.5~\si{\micro\metre}$  will make a lower instability threshold, this is because the impedance peak of the latter is much sharper as shown in Fig.~\ref{fig1}, which can also lead to MBI.  

\subsection{\label{secc}Impact of Bunch Lengthening with HHC}

\begin{figure}
\centering\includegraphics[width=8.5cm]{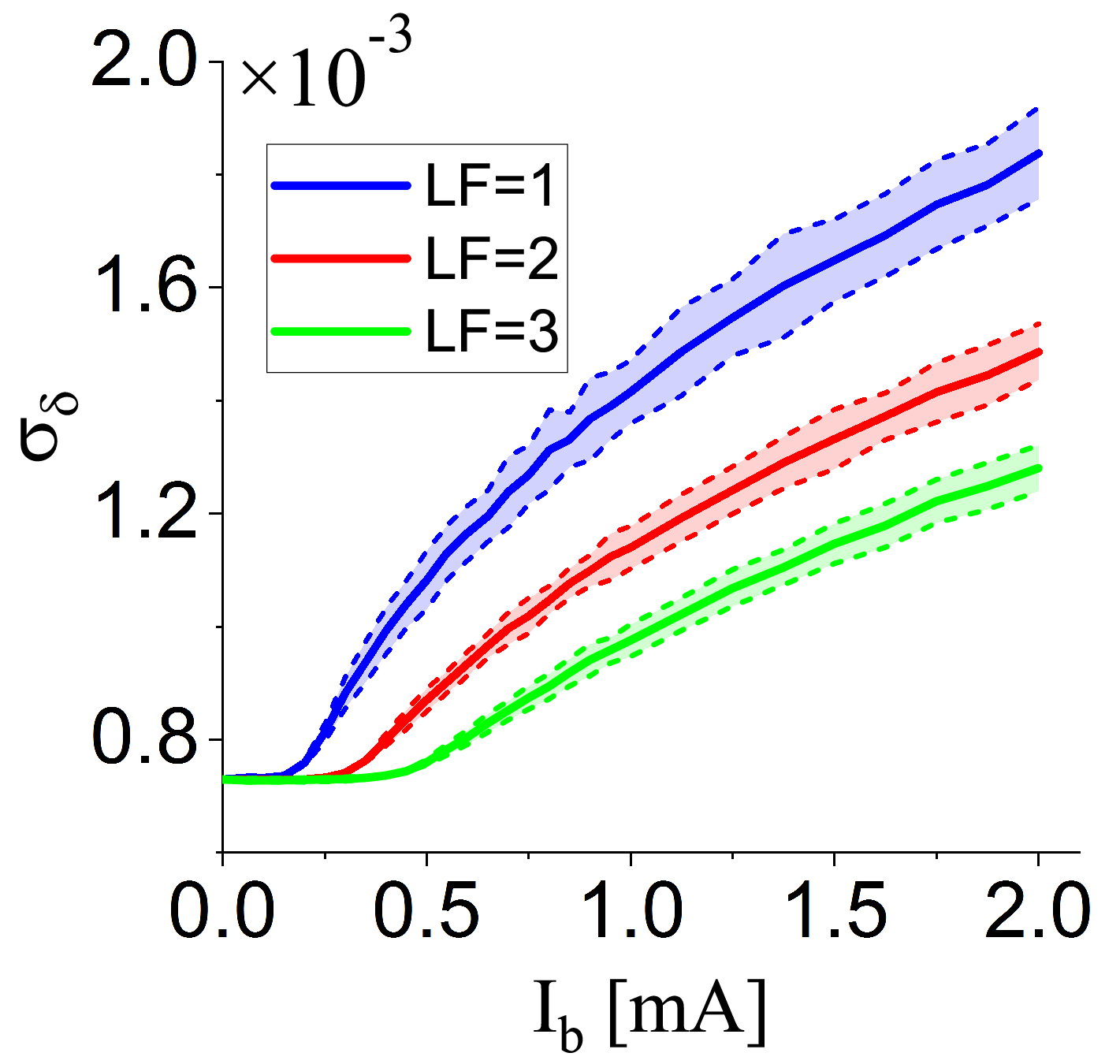}
\centering\includegraphics[width=8.5cm]{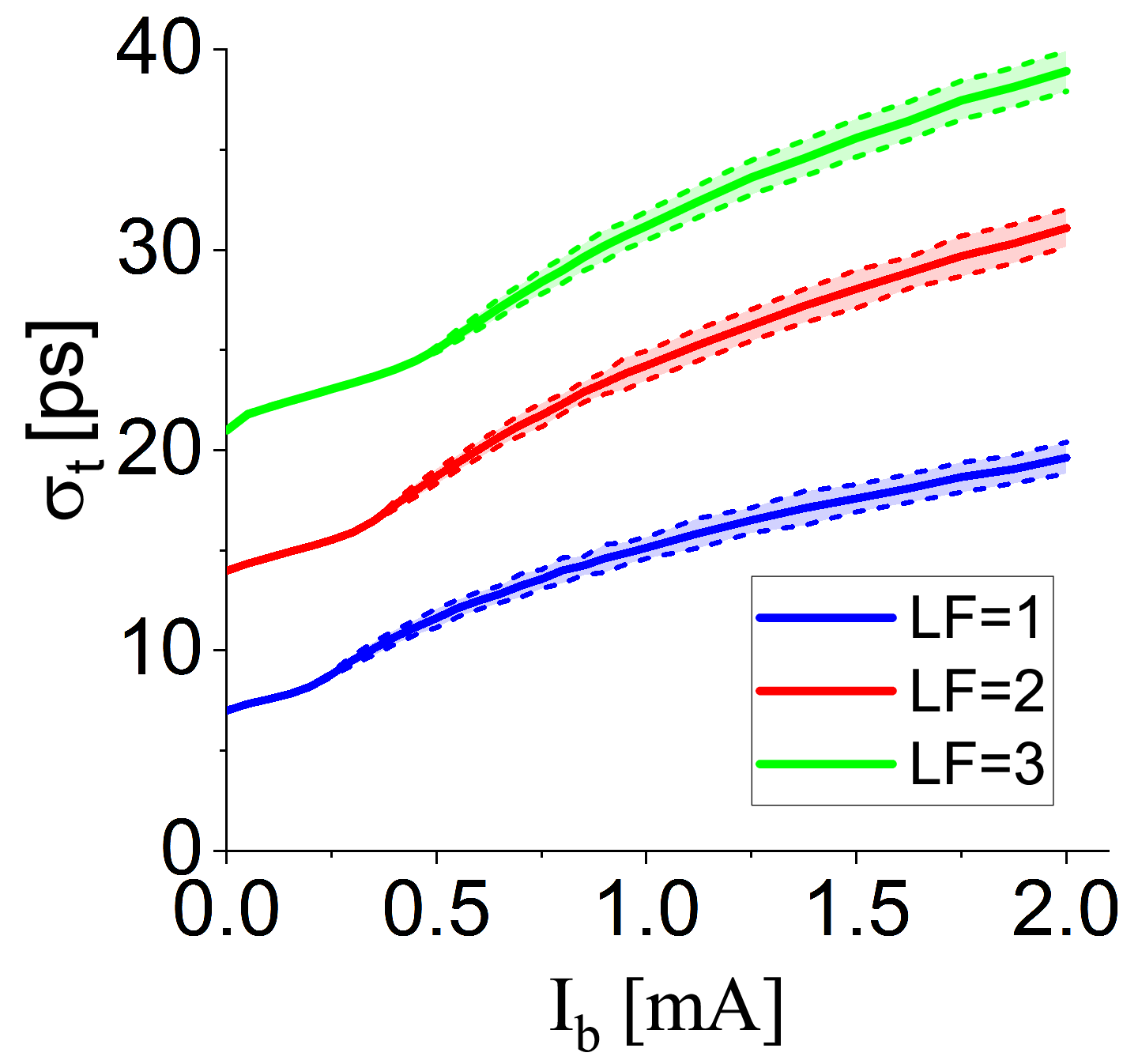}
\caption{The rms energy spread (top) and bunch length (bottom) versus single bunch current for $\rho_{\si{NEG}} =1\times10^{-5}~\si{\Omega~m}$ and $d=1~\si{\micro\metre}$  with different bunch lengthening factors (LFs). The solid lines are the mean values and the dashed lines including their fill areas represent the standard deviation.}
\label{fig9}
\end{figure}

Bunch lengthening with higher harmonic cavities (HHCs) is also a very helpful means to fight against most collective effects, including the MWI. A passive superconducting 3rd harmonic cavity  will be installed in the HALF storage ring \cite{He2021,He2022}. Instead of multibunch simulations, we just carry out single bunch simulations by introducing an ideal HHC voltage potential because of the heavy computational loads and  making the bunch length increase to a factor of 2 or 3 at zero current.  Figure~\ref{fig9} shows the predicted energy spread and bunch lengthening from the tracking simulations as a function of the single bunch current for the case of $\rho_{\si{NEG}} =1\times10^{-5}~\si{\Omega~m}$ and $d=1~\si{\micro\metre}$ with different bunch lengthening factors.  In order to obtain a convergent simulation, $\overline{\sigma}_s$ should still be as small as that without HHC to resolve the resonator-like peak impedance.  The bunch lengthening with HHC can  raise the MWI  threshold since it lowers  the charge density. 

\section{\label{sec5} Conclusion and Discussion}

In this paper, we have studied the impact of NEG coating resistive-wall impedance on the longitudinal microwave instability (MWI) for the HALF storage ring via particle  tracking simulation, where the wake potential of a very short Gaussian bunch with rms length of $\overline{\sigma}_s$ serves as  pseudo Green function. In order to obtain a quasi convergent simulation of the beam dynamics, $\overline{\sigma}_s$ should be small sufficiently to resolve the peak impedance in the high frequency region. For the cases presented in this paper, $\overline{\sigma}_s$ should be at most 0.02 mm, which is more than 100 times shorter than that of the equilibrium beam at zero limit. Otherwise, one is likely to  underestimate the MWI behavior. Recent microwave instability studies \cite{Carver2023} for the ESRF-EBS show that the measured current threshold is significantly lower than the stimulated one, but all the wakefield models are  computed with $\overline{\sigma}_s$ =1 mm, which is far from satisfying the resolution of the high frequency RW components and  can be a possible answer to explain the  discrepancy. 

The  effective impedance is often used to evaluate the MWI, however, our studies show that the  characteristics of the peak  in the high frequency region are also critical. A strong and narrowband peak can cause an undesirable micro-bunching instability (MBI), which has a low  threshold current and makes the dynamics of MWI more complex.  For a high coating resistivity  (in the order of $10^{-5} ~\si{\Omega~m} $), reducing the thickness is helpful to weaken the MBI by shifting the peak impedance to higher frequency but the MBI is still dangerous.  Another effective way to suppress the MBI is to reduce the coating resistivity, which has a broadband impedance.
The bunch lengthening with HHC can raise the MWI/MBI threshold. A storage ring with  NEG coating  of low resistivity applied to inner surfaces of many vacuum chambers and a low frequency main cavity \cite{Skripka2016b,brosi2023}  can still suffer from the MBI since it has high single bunch charges. 

The NEG-coating resistive-wall high frequency impedance can play an important role in the MBI, so accurate measurements of  the NEG  resistivity  are very important in order to  perform accurate simulations on their impact on the beam dynamics. The MBI will reduce the beam quality but also  has the potential to tailor the emitted CSR radiation and its fluctuations for possible applications of the terahertz radiation.

\begin{acknowledgments}
The authors would like to thank Sihui Wang at USTC and Na Wang at IHEP for useful discussions on NEG coatings and Biaobin Li at USTC for useful discussions on the numerical convergence. This work was supported by National Natural Science Foundation of China (No. 12105284 and No. 11875259) and the Fundamental Research Funds for the Central Universities (No. WK2310000090).
\end{acknowledgments}

\nocite{*}

\end{document}